\documentclass[modern]{aastex631}

\begin{document}

\shorttitle{Wilson-Harrington's Nucleus}
\shortauthors{Kareta and Reddy}

\title{Nuclear and Orbital Characterization of the Transition Object (4015) 107P/Wilson-Harrington}
\accepted{August 15, 2023}

\author[0000-0003-1008-7499]{Theodore Kareta}
\affiliation{Lowell Observatory, Flagstaff, Arizona, USA}

\author{Vishnu Reddy}
\affiliation{Lunar and Planetary Laboratory, University of Arizona, Tucson, Arizona, USA}
\begin{abstract}

Comet 107P/Wilson-Harrington, cross-listed as asteroid 4015, is one of the original transition objects whose properties do not neatly fit into a cometary or asteroidal origin. Discovered in a period of apparently gas-dominated activity in 1949, it was subsequently lost and recovered as the inactive asteroid 1979 VA. We obtained new and re-analyzed archival observations of the object, compared to meteorites, and conducted new orbital integrations in order to understand the nature of this object and to understand where it falls on the asteroid-comet continuum. Wilson-Harrington's reflectance spectrum is approximately neutral from visible to near-infrared wavelengths, but has a reflectance maximum near $0.8-0.9\mu{m}$. The object's spectrum is well matched by laboratory spectra of carbonaceous chondrite meteorites like the CM Murchison or the CI Ivuna. The object's phase curve slope is compatible with either an asteroidal or cometary origin, and its recent orbital history has no periods with high enough temperatures to have altered its surface. While it is possible that some unknown process has acted to change the surface from an originally cometary one, we instead prefer a fundamentally asteroidal origin for Wilson-Harrington which can explain its surface and orbital properties. However, this would require a way to maintain significant (hyper-)volatile supplies on the near-Earth objects beyond what is currently expected. Wilson-Harrington's similar metorite affinity and possible orbital link to sample return targets (162173) Ryugu and (101955) Bennu suggest that the returned samples from the Hayabusa-2 and OSIRIS-REx missions might hold the key to understanding this object.
\end{abstract}


\section{Introduction} \label{sec:intro}
\subsection{Historical Overview}
Despite a discovery more than seven decades ago, shockingly little is understood for certain about comet 107P/Wilson-Harrington. Discovered \textit{active} in November 1949 \citep{1949IAUC.1244....1B} as 1949g (modern naming C/1949 W1), it was subsequently lost due to a short observational arc facilitated by bad weather. Three decades later, \citep{1979IAUC.3422....1H} discovered the asteroidal (e.g., not showing signs of ongoing mass loss) 1979 VA in a comet-like orbit which \citet{1992IAUC.5585....1B} later propagated the orbit of back to 1949 -- and it matched that of Wilson-Harrington. (This results in the object's cross-listing as both comet 107P and asteroid 4015.) While Wilson-Harrington's activity might not have lasted the whole 1949 apparition, it has never been seen losing mass since its asteroidal recovery. Wilson-Harrington is thus one of the first members of the ``transition objects" -- objects whose properties do not fit cleanly into the ``inactive asteroid" or ``active comet" categories\footnote{In this paper, an object having a cometary nature or origin will typically be used to mean that the object has an Outer Solar System origin, as opposed to the object having spent most of its time in the Main Asteroid Belt.} -- along other compelling objects like (3200) Phaethon \citep{1983IAUC.3881....1W,2010AJ....140.1519J,2018AJ....156..287K} or 133P/Elst-Pizarro \citep{1996IAUC.6456....1E,2004AJ....127.2997H}.

\subsection{Activity in 1949 and Subsequent Searches}
\citet{1997Icar..128..114F} re-examined the original Palomar Observatory Sky Survey (POSS) 1949 discovery plates \citep{1949IAUC.1244....1B}. On the two plates where a tail can be seen, the tail was \textit{approximately} anti-sunward, not in the same direction as the streaking from the long sidereally-tracked exposures, very blue in color, and contemporary reports suggested it dissipated over at most a few days. The width of the point-spread function (PSF) of Wilson-Harrington in the cross-streak direction was found to be no wider than nearby field stars. If a coma was present, it must have been small with a scale height of a few hundred kilometers or less. The blue color and hours-to-days duration of the tail were interpreted by \citet{1997Icar..128..114F} to be only explainable as an ion tail of approximately equal amounts of $CO^+$ and $H_2O^+$, with implied water and carbon monoxide production rates of $\sim5\times10^{27}mols/s$ each as a ``plausible maximum." (Dust of the right size from \citet{1968ApJ...154..327F} modeling would not dissipate so quickly, and matching the blue color with smaller grains is tricky.) On those two plates alone, Wilson-Harrington looks to have an overall production rate on the lower end for typical comets, a significantly higher amount of $CO$ compared to $H_2O$ than typical comets observed at comparable heliocentric distances, and an apparently \textit{very} high gas to dust ratio.

Several attempts have been made to detect activity on the object in the decades after its initial 1949 outburst. \citet{1996Icar..119..173C} searched unsuccessfully for emission from the $CN$ radical, setting an upper limit on the $CN$ production rate an order of magnitude smaller than typical comets. \citet{1983Icar...54...59H} imaged 1979 VA and found their photometry somewhat noisier than expected, but still retrieved a rotation period near to the modern (somewhat uncertain) value. \citet{2003Icar..164..492L} specifically searched for a coma or tail and also came up empty-handed. For all intents and purposes, it appears that Wilson-Harrington has only been unambiguously detected as active for a few days in 1949 and never since.

\subsection{Big Picture Questions and Modern Hypotheses}
What is the actual origin of 107P/Wilson-Harrington: is it a dormant traditional comet derived from the Trans-Neptunian belt, or does it share more in common with the active asteroids derived from the Main Belt? (A recent review of the ``Asteroid-Comet Continuum" is provided in \citealt{2022arXiv220301397J}.) While \citet{1997Icar..128..114F}'s interpretation of the 1949 outburst as containing a very large amount of carbon monoxide -- a substance yet to be detected in any active asteroid or ``main belt comet" -- might seem like unambiguous evidence for a traditional comet origin, we will argue in this section that there are several properties of this object which fit the scenario less cleanly. 

While it possesses low visible and near-infrared albedos similar to that of comets ($p_J=0.05\pm0.01$ measured in the J filter, \citealt{1995P&SS...43..733C}, and $p_V=0.046\pm0.008$ measured in the V filter, \citealt{2011ApJ...731...53M}, and thus a diameter near $D\sim3.8km$\footnote{https://ui.adsabs.harvard.edu/abs/2016PDSS..247.....M/abstract}), its surface spectral properties as determined from broadband colors are not traditionally cometary. In the \citet{1984PhDT.........3T} taxonomy, Wilson-Harrington is a C or F type ``asteroid", indicating a nearly-neutral visible reflectance spectrum. A low albedo is also compatible with these kinds of asteroids, as well as other low-albedo types. TNO-derived comets generally have red-to-very-red nuclear spectra (see \citealt{2021PSJ.....2...31K} for a discussion of how reliable such a statement might be), while there are many neutral-reflectance asteroids spread throughout the Main Belt and Near-Earth space, including the majority of the active asteroids \citep{2022arXiv220301397J}. 

While Wilson-Harrington's Tisserand Parameter with respect to Jupiter is formally compatible with a cometary origin at $T_J = 3.08$, it is on the higher end (like 2P/Encke) and within the range of values where some contamination by Main Belt asteroids of the Jupiter Family Comets can be expected \citep{2020AJ....159..179H}. The Near-Earth Object (NEO) distribution model of \citet{2018Icar..312..181G}, accessed through the Lowell Observatory's \textit{astorb} database\footnote{https://asteroid.lowell.edu/astinfo/} \citep{2022A&C....4100661M}  gives Wilson-Harrington's orbit only a $7.7\%$ chance of originating from the JFCs, as opposed to $17.2\%$, $41.3\%$, and $32.4\%$ chances for the $\nu_6$, 3:1, and 5:2 complexes respectively. The model of \citet{2002Icar..156..399B} gives Wilson-Harrington just a $\sim4\%$ chance of coming from the JFCs. The 3:1 is particularly interesting due to its proximity to the New Polana and Eulalia families (specifically pointed out for Wilson-Harrington in \citealt{2002Icar..156..399B}), which in turn are the most likely sources for the OSIRIS-REx and Hayabusa 2 mission targets, (101955) Bennu and (162173) Ryugu. Spectral similarities between Wilson-Harrington and some Main Belt Asteroids, namely members of the Polana and Themis families, were presented in \citet{2015DPS....4710605C}. The Polana family was also recently proposed as a possible source of (3200) Phaethon as well \citep{2021Icar..36614535M}. (If Wilson-Harrington experienced significant non-gravitational acceleration in the past, as might be expected if it was more active or more frequently active at that time, then some of these orbital conclusions might be of limited utility.)

Carbon monoxide is significantly more volatile than water, so \citet{1997Icar..128..114F}'s interpretation of the 1949 activity as containing an approximately equal amount of the two well within the water ice line at $R_H=1.15$ AU is quite surprising. Not only do most comets have significantly higher water production rate compared to $CO$ at these kinds of heliocentric distances, one would expect that for a candidate ``dormant comet" like Wilson-Harrington that the $CO$ would have been depleted \textit{before} the water was. However, there is a clear counter-example to this first-principles argument: the discovery of $CO$-or-$CO_2$-driven activity in Near-Earth Object (3552) Don Quixote by \citet{2014ApJ...781...25M, 2020PSJ.....1...12M} at a heliocentric distance of $R_H=1.23$ AU. Don Quixote's orbit had long been suspected of having a cometary origin \citep{1985Icar...61..417H, 2002Icar..156..399B} but no activity had ever been found. The gas production rate of Don Quixote appears to scale with the inverse of its heliocentric distance squared, consistent with the ongoing sublimation of a hypervolatile ice. In other words, Don Quixote is another object with very limited dust production and significant hypervolatile production, but the sustained nature of Don Quixote's activity and it's classic cometary D-type reflectance spectrum stand in contrast with Wilson-Harrington's one-off activity and more neutral spectral behavior. Don Quixote ($T_J=2.32$) and Wilson-Harrington ($T_J=3.08$) have very different modern day orbits, and thus almost certainly very different orbital histories and dynamical stabilities, and thus have encountered different modern and historical thermal conditions that should render their ability to retain volatiles over the long-term different.

In summary, while 107P/Wilson-Harrington might have several properties which are similar to those of typical comets, it has just as many more reminiscent of the active asteroids or just downright puzzling instead. It thus seems clear that a re-inspection of Wilson-Harrington's properties and likely origin scenarios is well warranted by its potential importance to understanding the comet-asteroid continuum, how traditional comets go dormant, the volatile reservoirs of the active asteroids/main belt comets, and possibly even to interpreting the data obtained by the OSIRIS-REx and Hayabusa-2 spacecraft -- as well as their returned samples.

\section{Nuclear Properties} \label{sec:obs}
In this section, we first construct a combined visible-near-infrared reflectance spectrum of 107P/Wilson-Harrington using both new visible observations (spectroscopic and photometric) and archival near-infrared spectroscopy. We then use this and other additional information (nuclear albedo, phase curve, and phase reddening) to make detailed comaparisons of this object against comets, asteroids, and meteorites alike. A summary of all new and archival observations are available in Table \ref{tab:obs_log}.
\subsection{New Visible Photometry and Spectra}
\begin{figure}[ht!]
\plotone{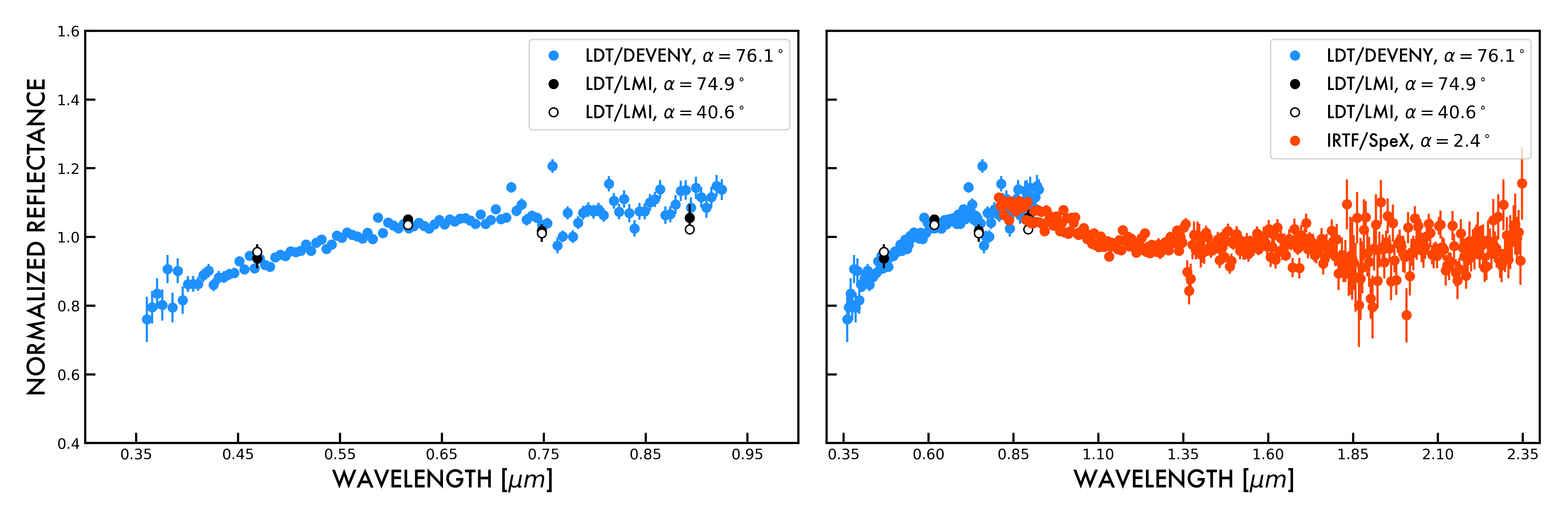}
\caption{The newly acquired visible (blue) and archival near-infrared (red) reflectance spectra of (4015) 107P/Wilson-Harrington are shown, along with two dates of complimentary visible photometry (black filled and unfilled circles), normalized to $R(\lambda)=1.0$ at $0.55\mu{m}$ and aligned such that their median values are identical over the $0.8-0.9\mu{m}$ wavelength range. The error bars on the visible photometry of Wilson-Harrington are only slightly larger than the data points. The two spectra agree well in curvature and slope in region where they overlap, though the visible spectrum is slightly redder likely due to phase-reddening.}
\label{fig:spectra}
\end{figure}

We first imaged Wilson-Harrington through the Sloan g, r, i, and z filters \citep{1996AJ....111.1748F} with the Large Monolthic Imager (LMI, \citealt{2014SPIE.9147E..2NB}) on the $4.3m$ Lowell Discovery Telescope (LDT) on 2022 May 31 (UTC) and 2022 September 5 (UTC). We interwove observations in g, i, and z with observations in the r filter to account for lightcurve-induced variations in the brightness of the object. After debiasing and flatfielding the images, we aligned, calibrated, and extracted the photometry of Wilson-Harrington using the \textit{PhotometryPipeline} package \citep{2017A&C....18...47M} using default settings for a point source. The brightness of each of the non-r filter observations was compared to the two nearest r filter observations, so effectively all colors were measured with respect to the r filter lightcurve. All non-r filter observations were bracketed, so no extrapolation of the r light curve was necessary for calculating the colors. The errors on the derived colors correspond to both the strict photometric errors from each individual frame and the uncertainty from the lightcurve correction together averaged over the number of utilized frames. The derived photometric colors were then converted to reflectivity values using the Solar colors of \citet{2018ApJS..236...47W} and normalized at $0.55\mu{m}$ by a linear interpolation between the g and r filters. The derived reflectance values for the May 31 and September 5 observations are shown as black unfilled and filled circles respectively in Figure \ref{fig:spectra} and listed in Table \ref{tab:colors}. We note that our colors are relatively good agreement with the colors of \citet{2011Icar..215...17U} after conversion between photometric systems.

Visible spectra were obtained of Wilson-Harrington on 2 September 2022 UTC with the DeVeny spectrograph  \citep{2014SPIE.9147E..2NB} on the LDT with the low-resolution ``DV1" grating and a 3 arcsecond wide slit providing a maximum spectral range of $\sim0.32-\sim1.0\mu{m}$. Conditions were good with great sky transparency, few or no clouds, and stable seeing. Twelve usable $300 s$ exposures were obtained for a combined one hour on source. Arc lamp observations were obtained at each pointing for wavelength calibration and flexure correction, and observations of a flatfield screen in the dome were used to correct for the illumination pattern. An identical observation scheme was used to observe the well-studied solar analog SAO 93936 (Hya 64) directly afterwards. We clipped all wavelengths short of $0.35\mu{m}$ and above $0.95\mu{m}$ for SNR reasons. The final constructed reflectance spectrum is thus the ratio of Wilson-Harrington's average spectrum divided by that of SAO 93936 after the application of an extinction correction to each individual spectrum.The data were reduced fully using the open-source spectroscopic data reduction package \textit{PypeIt} \citep{pypeit:joss_pub,pypeit:zenodo}. We note that the comet was observed over a wider range of airmass ($1.72 - 1.29$) than the solar analog was ($1.09-1.09$). We checked out retrieved spectrum against our Sep. 5 photometry and found that they agreed remarkably well, suggesting that our airmass correction scheme and good conditions produced a reliable spectrum. The final constructed visible reflectance spectrum of Wilson-Harrington is shown in blue in Figure \ref{fig:spectra}.

\startlongtable
\begin{deluxetable*}{c|c|c|c|c|c|c}
\tablecaption{Log of new and archival observations of 107P/Wilson-Harrington.}
    \label{tab:obs_log}
    \centering
    \startdata
        Tel./Inst./Config. & Date (YMD) & $R_H$ (AU) & Phase Angle & Airmass & $N_{exp}$ & $\tau_{exp}$ (s)\\
        \hline
        LDT/DeVeny/DV1 & 2022-09-02 & $0.974$ & $76.1$ & $1.72-1.29$ & $12$ & $300$ \\
        LDT/LMI/griz & 2022-05-31 & 1.465 & 40.6 & $\sim1.8$ & 16r, 3i, 3g, 3z & $12$ \\
        LDT/LMI/griz & 2022-09-05 & 0.980 & 74.9 & $\sim1.8$ & 40r, 3i, 6g, 6z & $20$ \\
        IRTF/SpeX/Prism & 2009-06-12 & 1.861 & $2.4$  & $1.01-1.01$ & $10$ & $120$
    \enddata
\end{deluxetable*}

\startlongtable
\begin{deluxetable*}{c|c|c|c|c|c}
\tablecaption{Derived Photometric Colors of 107P/Wilson-Harrington.}
    \label{tab:colors}
    \centering
    \startdata
        Target & Date (YMD) & Phase Angle & $g-r$ & $r-i$ & $r-z$\\
        \hline
         107P/W-H & 2022-09-02 & 76.1 & $0.51 \pm 0.02$ & $0.08 \pm 0.01$ & $0.13 \pm0.02$\\
         107P/W-H & 2022-05-31 & 40.6 & $0.48 \pm 0.01$ & $0.09 \pm 0.01$ & $0.14 \pm0.01$\\
        Sun (PS1) & - & - & $0.39$ & $0.12$ & $0.13$
    \enddata
\end{deluxetable*}

\subsection{Archival Near-Infrared Spectrum}
A near-infrared reflectance spectrum of Wilson-Harrington obtained on 12 June 2009 \citep{reddy2009} with the SpeX instrument \citep{2003PASP..115..362R} on the $3.0m$ NASA Infrared Telescope Facility (IRTF), which we accessed through the Planetary Data System's (PDS) Small Bodies Node (SBN). To understand the systematics of this dataset better and to try to improve upon its SNR, we retried the raw data files from the IRTF Legacy Archive\footnote{https://irtfdata.ifa.hawaii.edu/} and re-reduced them from scratch in \textit{spextool} \citep{2004PASP..116..362C}. The observation scheme ``book-ended" observations of Wilson-Harrington with observations of a near-by G-type star (SAO 185066) for correction of telluric absorption by the atmosphere. The proper solar analog SAO 120107 was also observed near $\sim1.0$ airmass to further correct the slope of the retrieved reflectance spectrum. All observations were obtained at or near to the parallactic angle to minimize the effect of atmospheric diffraction. After construction of a master 2D wavelength map and flatfield, we extracted each spectrum (comet and star) optimally using identical default settings for an unresolved point source. The telluric features in each comet spectrum were them corrected by division with the stellar observations directly before or after that block of comet observations and after all the comet observations were robustly combined, their slopes were corrected further by the difference in spectral slope and curvature of the standard star and the Solar Analog. All in all, 34 $120s$ exposures were obtained totaling 68 minutes on source. The final retrieved near-infrared reflectance spectrum of Wilson-Harrington is shown in red alongside the new visible spectrum in Figure \ref{fig:spectra}. As these data were obtained with SpeX before its 2014 upgrade (before which it had very low sensitivity below $0.8\mu{m}$) and as we saw very significant spectral variation from frame-to-frame at the shortest wavelengths in the comet, we cut all data below $0.8\mu{m}$ from the final spectrum as we could not be convinced of their reliability. We also cut all data beyond $2.35\mu{m}$ in wavelength as the SNR was dropping rapidly. As can be seen, the signal to noise on the spectrum is quite good and the spectral slope and behavior are easily compatible with the visible spectrum obtained over thirteen years later.

\subsection{Description of Combined Spectrum and Comparison to Other Comets}
The spectrum, as indicated by previous broadband observations and the visible/near-infrared albedo constraints \citep{2011ApJ...731...53M, 1995P&SS...43..733C}, is approximately neutral over the whole wavelength range studied. A red spectrum short of $\sim0.55-0.60\mu{m}$ shallows and then becomes weakly-blue sloped by $\sim0.75-0.85\mu{m}$, again becoming neutral beyond $\sim1.1\mu{m}$. The spectral curvature and slope changes must be real and not artifacts of the reduction process: they show up in individual frames from each instrument and all data were processed and analyzed by packages and scripts which had been used successfully for asteroid and comet observations in the past.

We note that the visible reflectance spectrum obtained with LDT/DeVeny \citep{2014SPIE.9147E..2NB} in September agrees well with the LDT/LMI photometry from three days later but is slightly redder than the reflectivity derived from photometry from May. Indeed, the reflectance photometry from May agrees even better with the near-infrared spectrum at the overlap wavelengths. We strongly suspect the primary reason for this is \textit{phase-reddening}, whereby objects viewed at higher phase (Sun-target-observer) angles can appear ``redder" than those observed at lower phase angles. (See \citealt{2012Icar..220...36S} for more information.) In essence, the photometry-derived reflectivity -- approximately neutral over the visible wavelength range with a slight reddening shortwards of $0.55\mu{m}$ -- we infer to be closer to how Wilson-Harrington reflects light at low phase angles. Given that phase reddening seems to be most important at shorter wavelengths for asteroids \citep{2012Icar..220...36S} and comets \citep{2015A&A...583A..30F}, we suspect that the near-infrared spectrum is accurate as-is.

We thus re-summarize Wilson-Harrington's reflective properties at low phase angles where they are most easily compared to other objects and laboratory samples: Wilson-Harringon has a reflectance spectrum that is approximately neutral over the $0.35-2.4\mu{m}$ wavelength range, with a slight decrease in reflectivity at short wavelengths and a change from a neutral-red slope shortwards of $\sim0.75-0.85\mu{m}$ becoming blue-sloped until $\sim1.1\mu{m}$ after which it becomes spectrally neutral again. 

The reflectance spectrum of Wilson-Harrington's nucleus looks very different from comet nuclei, which are traditionally thought to be very red throughout this whole wavelength range like the D-type asteroids. This was already apparent from the visual colors of Wilson-Harrington (see, e.g., \citealt{1984PhDT.........3T}) being anything other than ``very red" and the similar albedos at visible \citep{2011ApJ...731...53M} and near-infrared \citep{1995P&SS...43..733C} wavelengths, but our new combined reflectance spectrum adds several new pieces of information to the puzzle. While cometary nuclei having slope changes in the near-infrared is common (somewhere between $0.8-1.25\mu{m}$, \citealt{2021PSJ.....2...31K}), they are typically slight decreases in slope from ``very red" down to ``red." (The D-type asteroid end-member in the Bus-Demeo taxonomy, \citealt{2009Icar..202..160D}, also shows a subtle blue-ing at these wavelength ranges.) \citet{2021PSJ.....2...31K} suggested that the surface of almost-certainly-a-dormant-comet (196256) 2003 EH$_1$ could be explained as a \textit{thermally altered} version of a traditional comet surface. In essence, close passages near the Sun can induce chemical changes that can change the reflectance spectrum of the comet nucleus or asteroid surface -- often assumed to make the surface more spectrally ``blue" based on laboratory data. (\citealt{2022PSJ.....3..187H} even recently found that near-Sun asteroids are bluer than their cooler counterparts.) However, Wilson-Harrington's spectrum does not show the same spectral curvature as that formerly sungrazing dormant comet and it is challenging to assess the likelihood of this possibility without short-term high-resolution orbital integrations. This possibility and the neccessary integrations are discussed further in Section \ref{sec:orbit}. In summary, if Wilson-Harrington has a traditional cometary origin, it has a surface that reflects light unlike other known comets.

\subsection{Phase Curves from ATLAS}
Another way to compare the surface of Wilson-Harrington to other potentially similar objects would be to derive a photometric phase curve for it, which is a measure of how the object dims with increasing phase angle and is a function of surface texture and composition. We utilized the Asteroid Terrestrial-impact Last Alert System (ATLAS, \citealt{2018PASP..130f4505T}) ``forced photometry server"\footnote{https://fallingstar-data.com/forcedphot/} \citep{2021TNSAN...7....1S} to gather the photometric data for this analysis. This service allows users to query the ATLAS survey's database for individual objects and to obtain measurements of their brightness in the two ATLAS filters ($o$ for orange and $c$ for cyan) over some queried date range. The details of the ATLAS photometric calibration pipeline are available in \citet{2018AJ....156..241H} and \citet{2020PASP..132h5002S}.

Wilson-Harrington has only had one good apparition since the dawn of ATLAS, so we only utilized data after Modified Julian Date $59600$, or 2022 January 21. All observed magnitudes that were more than $3-\sigma$ above the noise floor and were dimmer than tenth magnitude (i.e. they were not contaminated by a bright field star) were corrected for their geometry ($\Delta$ and $R_H$) to phase-dependent absolute magnitudes, or $H(1,1,\alpha)$. This apparition provided good phase angle coverage between $\alpha\sim30^{\circ}$ and $\sim80^{\circ}$. This is a wider phase-angle range than was utilized in \citet{2011ApJ...726..101I} ($\alpha\sim39-68^{\circ}$). This does not include low enough phase angles to constrain any kind of opposition effect, so we restricted ourselves to a simple straight-line fit to the equation $H(1,1,\alpha) = H(1,1,0) - \beta \times \alpha$. $H(1,1,0)$ is the ``true" absolute magnitude (assuming the phase curve is completely linear and without an opposition surge, which is probably unrealistic) and $\beta$ is the linear phase function coefficient. We chose this formalism to explicitly compare to \citet{2017MNRAS.471.2974K}'s of comet nuclei properties. We also note that for most of the comets in the sample of \citet{2017MNRAS.471.2974K}, the phase angle range utilized was significantly narrower (typically $<10^{\circ}$) than what we were able to use here, so while the techniques employed are very similar, it is not necessarily a one-to-one comparison.

The data ($H(1,1,\alpha$) vs. $\alpha$) appeared as perfectly straight lines with scatter in both the $o$ and $c$ filters and we thus fit the two lines using the \textit{emcee} package \citep{2013PASP..125..306F}. While using Markov-Chain Monte Carlo (MCMC) techniques to fit a line might be viewed as overkill, the dataset contained a number of points that \textit{looked} like outliers and thus at least a Maximum Likelihood Estimation (MLE) approach was preferred over a traditional least-squares fitting routine. The retrieved corner plots showed perfectly gaussian contours and the expected covariance trends (e.g. a brighter absolute magnitude made for a steeper linear phase function coefficient).

We retrieved the absolute magnitudes $H_o = 15.75\pm0.05$ and $H_c=16.20\pm0.04$ and the linear phase function coefficients $\beta_o = 0.0427\pm0.0008$ and $\beta_c = 0.0401\pm0.0006$ measured in magnitudes dimmed per degree of phase angle. All reported values are the median value of all chains after the burn-in with the $68\%$ confidence intervals, similar to a $1-\sigma$ error, reported as the error. This is in good agreement with \citet{2011ApJ...726..101I}'s analysis, which found $\beta_R = 0.0410\pm0.002$ measured over a smaller range of phase angles. While the absolute magnitudes in each filter are very clearly different, the phase curves measured overlap at less than a $2-\sigma$ level and thus cannot really be distinguished within this ATLAS dataset. The two ATLAS filters, $o$ and $c$, have enough spectral overlap that measuring a phase reddening effect directly from these data to compare with our spectroscopic/photometric phase reddening interpetation in the previous sub-section was not possible. In other words, searches with filters more widely separated in wavelength or a dedicated spectroscopy campaign might be better suited to searching for phase-reddening more directly.

\begin{figure}[ht!]
\plotone{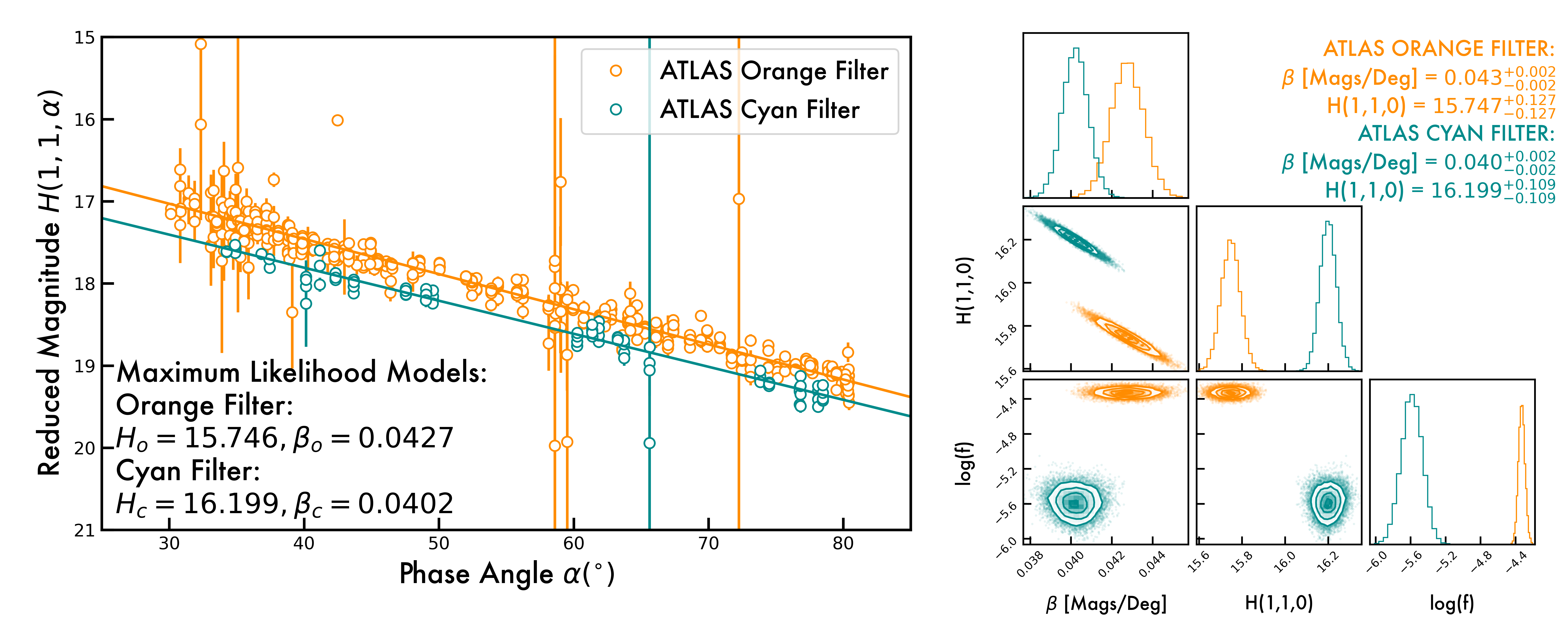}
\caption{Left: Observations of 107P/Wilson-Harrington by the ATLAS Survey \citep{2018PASP..130f4505T} through their Orange and Cyan filters are converted are corrected by the object's heliocentric and geocentric distances and compared against the maximum-likelihood parameter combination models from our MCMC analysis described in detail in the text. Right: the outcome of our MCMC analysis is shown as a corner plot, and the $99.7\%$ confidence intervals for both filters are listed in the top right.
As expected, the data and simple straight-line model ($H(1,1,\alpha) = H(1,1,0) + \beta \times \alpha$) agree well as we do not measure the object's brightness at small enough phase angles to probe any possible opposition surge. The slope of either curve is fully compatible with comets or low-albedo asteroids.}
\label{fig:atlas}
\end{figure}

While a wide variety of linear phase curve coefficients are allowed for asteroids, only a small range of values has been seen on comet nuclei. \citet{2017MNRAS.471.2974K} found a median linear phase curve coefficient of $\beta=0.042$ with a standard deviation of $0.17$ in their survey of comet nuclei, which is indistinguishable from either of our values. \citet{2017MNRAS.471.2974K} also found evidence for an albedo-$\beta$ relationship on comet nuclei, with nuclei with higher albedos also having steeper phase curves. Wilson-Harrington would plot perfectly along that trend, near 103P/Hartley 2 and 2P/Encke nearby. However, we note explicitly that some low-albedo asteroids have very similar linear phase curve slopes -- \citet{2013Icar..226..663H} found that OSIRIS-REx mission target (101955) Bennu had a phase curve slope of $\beta=0.040\pm0.003$ over the phase angle range of $\sim15^{\circ}-95^{\circ}$. It seems that Wilson-Harrington's phase curve coefficient is easily compatible with other comets or with asteroids that it might be dynamically linked to. Wilson-Harrington's thermal inertia and albedo are both also compatible with a possible cometary or asteroidal nature \citep{2009A&A...507.1667L, 2017AJ....154..202B}, highlighting the challenges with clearly discerning the origin of low-albedo objects generally -- a topic we return to in Section \ref{sec:origins}.

\subsection{Comparison to Meteorites and Asteroids}
While this might be an extremely interesting spectrum of a cometary nucleus, a near-Earth asteroid with these reflective properties might be only fairly interesting. In the Bus-Demeo taxonomy, there are no end-member classes or sub-classes which make the full switch from neutral-red slopes at visible wavelengths to blue/neutral slopes at near-infrared ones. However, the transition from blue to neutral in the near-infrared is reminiscent of the bluest asteroids like (3200) Phaethon \citep{2018AJ....156..287K} or to a much lesser extent (101955) Bennu \citep{2011Icar..216..462C,2019NatAs...3..332H}.
\citet{2010JGRE..115.6005C} finds many B-type asteroids which do the \textit{opposite} going from blue to red with increasing wavelength. If one surmises that at very low phase angles that Wilson-Harrington might have a neutral or neutral-blue reflectance spectrum at visible wavelengths, then some of the B-type asteroids in \citet{2012Icar..218..196D} might share some similar properties -- namely their ``G4" group associated with thermally altered CM Chondrites or perhaps their ``G6" group associated again with (3200) Phaethon, though neither match is satisfactory. None of the recent or near-future spacecraft targets provide good matches either, (162173) Ryugu \citep{2019Sci...364..272K, 2013Icar..224...24M} and 67P/Churyumov-Gerasimenko \citep{2016MNRAS.462S.476R, 2020NatAs...4..500R} are too red and (101955) Bennu \citep{2020A&A...644A.148S} and (3200) Phaethon \citep{2018AJ....156..287K} are too blue.

\begin{figure}[ht!]
\plotone{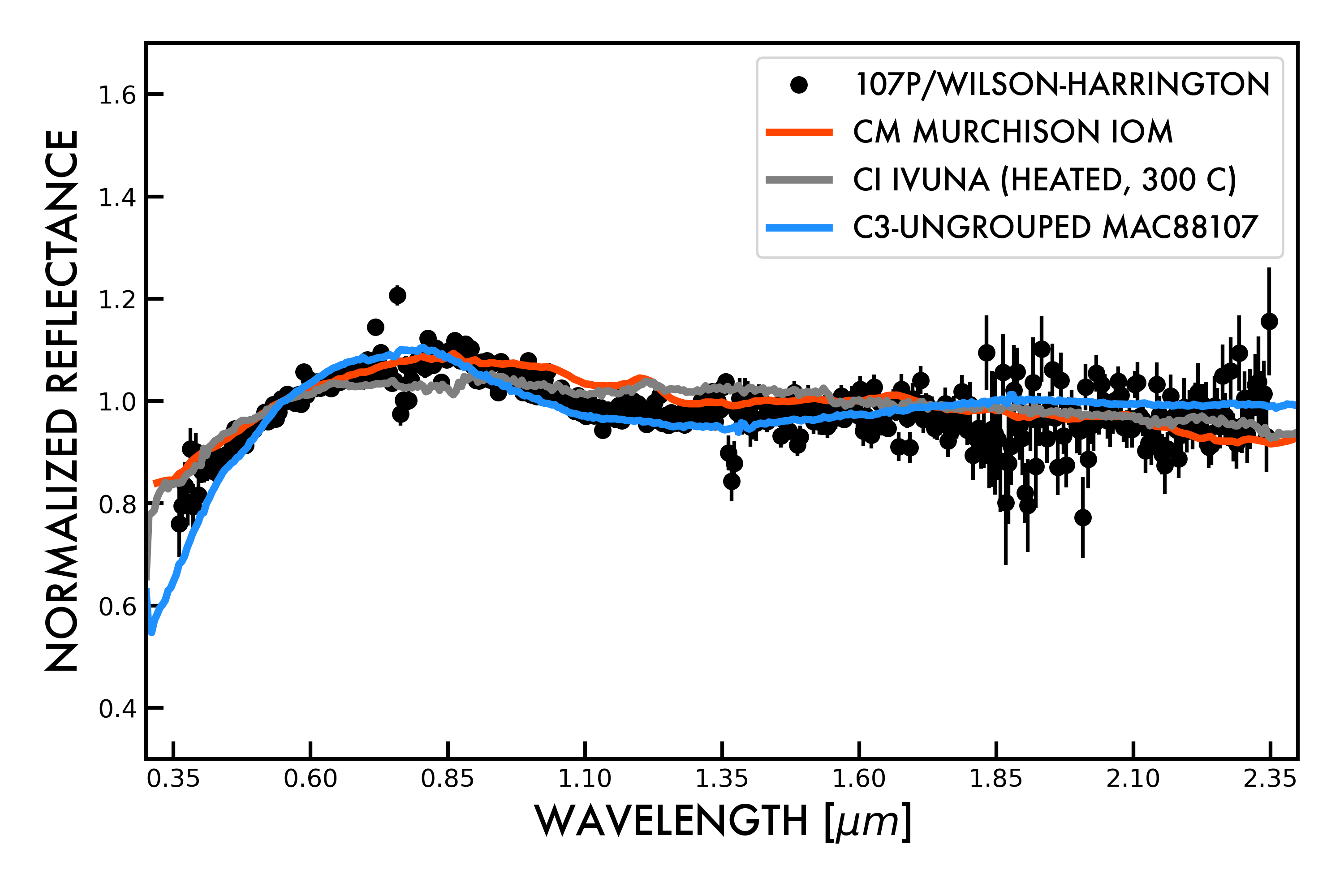}
\caption{The combined visible/near-infrared reflectance spectrum of 107P/Wilson-Harrington (black) is compared against the closest spectral matches in the RELAB database \citep{1983JGR....88.9534P}, the insoluble organic material (IOM) from the CM Chondrite Murchison (orange), a sample of the CI Ivuna after it had been heated to $300^{\circ}$ Celsius (grey), and the Ungrouped C3 chondrite MAC 88107 (blue). Murchison's IOM is the best fit spectrally, while MAC 88107 is the best match in albedo.}
\label{fig:comparison}
\end{figure}

We thus compared our combined visible/near-infrared reflectance spectrum of Wilson-Harrington against every reflectance spectrum in the RELAB database \citep{1983JGR....88.9534P} that had a sufficiently low albedo (we used a generous cut of $p_V < 0.15$) to be a plausible match to Wilson-Harrington's albedo ($p_V \sim 0.046$, \citealt{2011ApJ...731...53M}). In Figure \ref{fig:comparison}, we display our Wilson-Harrington spectrum against the three\footnote{The mineral Ilmenite, $FeTiO_3$, was also a surprisingly good match, but we restricted our analysis to meteorites and other astromaterials.} best matches : Insoluble Organic Material IOM) from the CM Chondrite Murchison\citep{1970Metic...5...85.} ($90-106\mu{m}$), a sample of the CI Chondrite Ivuna which had been heated to $300^{\circ}$ Celsius \citep{1996LPI....27..551H} ($<125\mu{m}$), and the ungrouped C3 Chondrite MAC 88107\citep{1994Metic..29..100G} ($<125\mu{m}$). All three laboratory spectra are good spectral matches and have low albedos -- Murchison's IOM has  $p_V\sim0.010$, the heated Ivuna sample has $p_V\sim0.023$, and MAC 88107 has $p_V\sim0.050$.

There are no known meteoritical analogues to traditional very-red spectra of cometary nuclei, but we have found several matches to our reflectance spectrum of Wilson-Harrington among the Carbonaceous Chondrites (CCs). MAC 88107 is petrologic type three, Murchison type two, and Ivuna type one -- while meteorites of type three have barely been aqueously altered, those of lower-numbered types are increasingly so -- and thus our best matches span the range of aqueous alteration states commonly found on the CCs. While these laboratory spectra might be from somewhat rare meteorites (like the CIs) or from sub-samples of larger rocks (the IOM in Murchison), the strangeness of our matches might have been expected from the apparent rarity of other objects with Wilson-Harrington-like spectra. We remind the reader that the visible spectrum is likely phase-reddened slightly, making the slope at visible wavelengths likely better fit by the Ivuna and Murchison samples as opposed to the spectrum of MAC 88107. We discuss the possible importance of these matches in Section \ref{sec:origins}.

\section{Orbital Evolution and History} \label{sec:orbit}
As mentioned in Section \ref{sec:intro} and subsequently, the \citet{2018Icar..312..181G} model for the distributon of near-Earth object orbits and origins does not place high confidence in the Jupiter Family Comet origin of comet 107P/Wilson-Harrington -- just a $7.7\%$ chance according to the Lowell Observatory \textit{astorb} database \citep{2022A&C....4100661M} compared to $\sim41.3\%$ of coming from the 3:1 complex and thus a possible link to the (New) Polana or Eulalia familes (and thus a compositional and origin link to Ryugu and Bennu and possibly Phaethon).

\begin{figure}[ht!]
\plotone{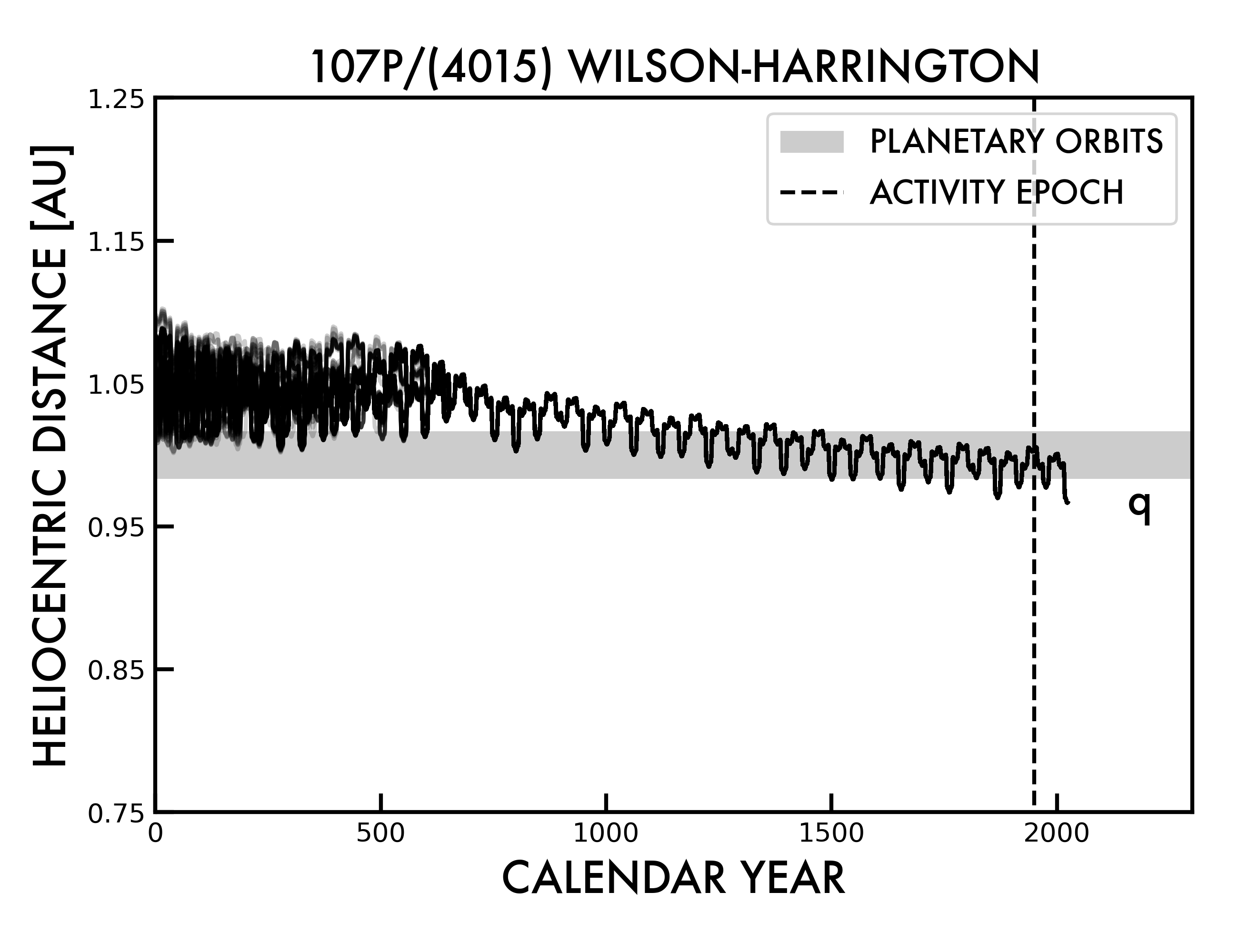}
\caption{The evolution of (4015) 107P/Wilson-Harrington's nominal orbit from late 2022 backwards 2000 years in time, as well as the identical propagation of 1000 clones sampled from the covariance matrix of its orbital fit (JPL 282). Despite Wilson-Harrington's very near-Earth perihelion and low inclination, its orbit is remarkably stable over the time period -- including no phases as a near-Sun object like some other dormant comets that might correlate with altered surface reflective properties from intense heating \citep{2021PSJ.....2...31K}. Integrations further backwards in time continued to chaotically diverge but showed no obvious trends or close encounters.}
\label{fig:orbit}
\end{figure}

To follow up on the previous section's suggestion that Wilson-Harrington's reflectance spectrum resembles some heated carbonaceous chondrites, we investigated the recent orbital history of the object using the \textit{Rebound} orbital dynamics package \citep{2012A&A...537A.128R}. We integrated the nominal orbit of Wilson-Harrington as well as that of 1000 clones sampled from the covariance matrix of the orbit solution (JPL 282) back four thousand years from the solution epoch. The IAS15 integrator \citep{2015MNRAS.446.1424R} was utilized for the integrations as it has an adaptive timestep critical for properly resolving close encounters between the test particles and the massive bodies (the Sun and the eight planets). Close encounters were expected in Wilson-Harrington's orbital history -- it has a relatively short orbital period for a comet ($P = 4.25$ years) and a low orbital inclination of just $2.8^{\circ}$ at present. The four-thousand year integration period was selected to encompass the likely period over which back-integrating the object's orbit might be deterministic and physically meaningful. The results of the integrations -- the perihelion distance of Wilson-Harrington over the most recent two-thousand year subset of the integration interval during which the orbital clones are only starting to diverge -- are shown in Figure \ref{fig:orbit}.

Surprisingly, Wilson-Harrington's orbit has changed very little over the past four thousand years. For at least the past fifteen hundred years, the trend has been to slowly decrease its perihelion distance a fraction of an AU per century modulated by short-term periodic variations caused by three-body interactions. One immediate conclusion from even a cursory glance at Figure \ref{fig:orbit} is that Wilson-Harrington has not been in an orbit hot enough to induce thermal alteration \citep{2021PSJ.....2...31K} let alone a properly sungrazing one recently. Even if it had been, it is not clear how its 1949 activity would have maintained or erased such a surface signature. The association of thermally-altered meteorite spectra with objects that are later found out to be comparatively more hydrated than their meteoritical analogues happened for both Bennu and Ryugu \citep{2019NatAs...3..332H, 2019Sci...364..272K}. In other words, while these objects clearly still have properties of the meteorites they've been linked to \citep{2022SciA....8D8141H}, actually inferring the level of heating that happened to the object seems to be less reliable. Taken as a whole, this makes the possible match to a ``lightly" heated sample of Ivuna described in the previous section less preferred compared to the other possibile matches. This does not rule out some even-further-back period of significant heating that might have altered the surface of the object, but we cannot assess the likelihood of this with backwards integrations.

The second and perhaps more crucial insight into Wilson-Harrington's history in our integrations is that Wilson-Harrington is \textit{slightly} warmer now than it was in 1949, and thus that any volatile substances would have a slightly easier time sublimating now than they would have in its one epoch of detected activity. Indeed, it has been warmer than its activity epoch (labelled in Figure \ref{fig:orbit} with a vertical line) for much of the time since -- including when it was re(dis)covered in 1979. (Granted, this difference is at most a few Kelvin.) While the actual process that drove Wilson-Harrington's 1949 outburst or increase in activity is not clear, if it was thermal in nature (e.g., driven by incoming solar insolation), it would be \textit{easier} for that process to proceed now and in several epochs in the recent past. The fact that this object has never been seen active since places a few key constraints on the problem.

The first is that whatever process caused Wilson-Harrington to release material in 1949 was most likely a rare sporadic event not related to how warm it was or where it was in its orbit. This could be solar activity induced as suggested by \citet{1997Icar..128..114F}, deeply buried ice suddenly sublimating vigorously as has been seen on otherwise-inactive objects like P/2010 H2 (Vales) \citep{2020PSJ.....1...77J}, or a small-but-significant impact that aids another process. The lack of any easy way to interpret the 1949 observations as being dusty is a truly puzzling issue -- it makes active-asteroid style explanations of activity (e.g, rotationally-driven mass loss or electrostatic lofting, both of which would be dust-dominated) more challenging to fit.

The second constraint is that Wilson-Harrington has been in the inner Solar System for at least a few thousand years if the lack of non-gravitational effects is a justified assumption. While ices can be held onto for a few-to-ten thousand years \citep{2010AJ....140.1519J} at depth on kilometer-scale near-Earth objects, this is considerably more challenging for more volatile substances (carbon monoxide in this case) and is related to the rotational state and obliquity of the object. Wilson-Harrington's outer surface and at least some fraction of its interior is likely equilibrated with its modern orbital/thermal conditions.

Where and for how long ice can survive on asteroids is a topic of considerable ongoing research, but obliquity, rotation state, and upper regolith structure all play significant roles (see, e.g., \citealt{2008ApJ...682..697S,2009MNRAS.399L..79P}). In general, the shallower the diurnal, seasonal, and orbital thermal waves can propogate, the easier it is to retain ice close to the surface. This means that low-obliquity asteroids, asteroids with highly insulating outer layers, and asteroids further from the Sun all can retain more volatiles for longer. Clearly some asteroids can preserve water ice for at least the age of the Solar System, so the question is how easily other kinds of ices were maintained and how long each kind of ice might persist after entrance to the NEO region.

\section{Origins} \label{sec:origins}
There is no obvious way to combine all of 107P/Wilson-Harrington's properties into a clean story. If the object had never been seen active -- or if the conclusion of \citet{1997Icar..128..114F} that the activity was gas-dominated is in some way incorrect -- the modern orbital properties and reflectivity of Wilson-Harrington could be linked to an origin in the Polana or Eulalia families easily. If the object had a more traditionally cometary reflectance spectrum, we could easily classify it as an analog of the nearly-dormant-but-gas-rich-comet Don Quixote \citep{2014ApJ...781...25M}. If it had an orbit that was cooling instead of warming slowly, we might be able to understand why its activity shut off -- the list goes on! We thus outline two broad scenarios that might explain most of the story and how to test aspects of them and move the study of the asteroid-comet-continuum forwards.

\subsection{Scenario 1: Unknown Cometary Surface Alteration Process}
If Wilson-Harrington is to have a traditional comet origin, we need to envision a process that can alter originally red and organic rich spectra into more spectrally neutral alternatives like what the object currently has. This process cannot be driven by close passages to the Sun and would need to act at larger heliocentric distances to reconcile its modern properties with its orbital and activity history. If we assume that Wilson-Harrington originally had a traditional comet reflectance spectrum, then there are many objects which are currently classified as asteroids which may actually trace their origins back to the TNOs rather recently. These would likely be low-albedo asteroids with red reflectance spectra (but not as red as comet nuclei), similar phase curves and thermal inertias \citep{2017AJ....154..202B} to comets and low-albedo asteroids, and are in orbits that are not exclusively cometary (e.g. $T_J$ around or slightly above $\sim3.0$). The general agreement between NEO origin models and the known NEOs about what fraction of the population might be dormant comets (a few percent, \citealt{2002Icar..156..399B, 2018Icar..312..181G}) suggests that these hidden comets shouldn't be too large in number. Furthermore, while \citet{2017MNRAS.471.2974K} inferred a relationship between changing surface properties (albedo, phase curve slope) and age, there is no established trend in nuclear colors for cometary nuclei with age. In other words, it would be challenging to explain Wilson-Harrington's spectrum as simply being an old and devolatilized comet -- some other process would have to be active or have been active. A third complicating factor is that Wilson-Harrington clearly lost \textit{some} mass in 1949, so this process would have to be resilient against or dominant over significant surface alteration of the object that comes with activity. (Maybe the comet's active fraction was so low this doesn't matter?) In general, while we are skeptical of invoking some yet-understood process to explain our spectrum of Wilson-Harrington, we cannot rule this scenario out completely.

\subsection{Scenario 2: Volatile Retention in Carbonaceous Asteroids}
For the comets currently recognized to have non-standard nuclear reflectance spectra, there is usually one of two explanations that can be utilized \citep{2021PSJ.....2...31K}. Either they aren't actually from the trans-Neptunian belt and are asteroidal interlopers (see, e.g., \citealt{2020AJ....159..179H}) or their orbital evolution has brought them close enough to the Sun for the intense heat to alter their surface properties \citep{2021PSJ.....2..190K, 2022PSJ.....3..187H}. This latter scenario does not appear to be the case for Wilson-Harrington in the past few thousand years, and the former scenario is already raised by other non-spectral lines of evidence. An origin in a carbonaceous main belt asteroid family \citet{2015DPS....4710605C} could explain its orbit \citep{2002Icar..156..399B,2018Icar..312..181G}, its phase curve \citep{2013Icar..226..663H, 2011ApJ...726..101I}, and a non-recurring activity would make it more similar to the active asteroids than the traditional comets \citep{2022arXiv220301397J}. We note that this is the first work we are aware of that links a designated comet with a meteorite analog -- if Wilson-Harrington is not a carbonaceous asteroid interloping among the JFCs, then one needs to consider why this single comet is the only one for which this process has been successful.

If Wilson-Harrington's origin lies in the Main Belt, the question then becomes how it was able to maintain a supply of volatiles -- not just water ice, but at least a significant amount of carbon monoxide if \citet{1997Icar..128..114F}'s conclusions are correct -- over thousands of years in near-Earth space. While preserving water ice over that time period might be plausible, preserving any carbon monoxide even at Main Belt temperatures would be comparatively more challenging. However, \citet{2006Icar..182..161L} suggested one way of forming the (similarly short) orbit of comet 2P/Encke -- slightly larger to Wilson-Harrington -- was a long period of storage in the Main Belt -- and while depleted, carbon monoxide has been clearly detected in that object \citep{2018AJ....156..251R}. The origin of Encke's orbit remains a topic of research, but at the very least the detection of CO at Encke suggests that even in environments much warmer than what Wilson-Harrington experiences, \textit{some} hypervolatiles can be preserved for significant periods of time. (It is worth stating explicitly that, again, these two objects have significant differences in their current and historical orbits, so some differences are to be expected.) While the specific outcomes of different thermal models can predict radically different ice depths -- \citet{2009MNRAS.399L..79P} and \citet{2016Icar..276...88S} disagree about the possible depth of ice on 133P/Elst-Pizarro, which shows recurrent cometary activity near perihelion and is usually considered the first-recognized Active Asteroid, by about a factor of $\sim50\times$, \citet{2020Icar..34813865S} showed that for low-obliquity NEOs that originated in the Outer Main Belt, ice can be easily preserved in the polar regions through the most common dynamical pathways to near-Earth space. Perhaps if everything goes right for a given NEO -- low spin obliquity, low surface thermal inertia, and a relatively recent exit from the Outer Main Belt -- species more volatile than water could be retained.

While the prospect of some of the carbonaceous asteroids retaining carbon monoxide over four-and-a-half billion years in the Main Belt might have significant theoretical hurdles to cross, this scenario is at least easier to investigate with the recent return of samples from Ryugu via Hayabusa-2 and the upcoming return of samples from Bennu via OSIRIS-REx. Indeed, some of the first analysis of the Ryugu returned samples was of extremely tiny fluid inclusions inside of the grains themselves \citep{2022LPICo2695.6011Z}. Perhaps the material seen leaving Wilson-Harrington in 1949 was not being produced by sublimation but instead was trapped inside of grains and interstitially inside of minerals instead. The analysis of the Ryugu and Bennu samples over the coming years and decades might then help us interpret the observations made of this object in the late 1940s.

\section{Summary}
Comet 107P/Wilson-Harrington, cross listed as asteroid 4015, is one of the original ``transition objects" whose properties do not neatly fit into the ``asteroid" or ``comet" categories. Wilson-Harrington was seen to be active for a few days in 1949 \citep{1997Icar..128..114F}, displaying a nearly-sunward and very blue tail and no clearly detected coma, was subsequently lost due to a short observational arc, and was then recovered as the apparently inactive asteroid 1979 VA. The incredibly blue tail was interpreted by \citet{1997Icar..128..114F} to be an ion tail composed of equal parts ionized water and carbon monoxide -- a high $CO/H_2O$ ratio for an active comet, made only more perplexing by the total lack of dust and short duration of its activity. While it has an orbit compatible with a cometary origin, it is equally if not more compatible with an origin in the Main Belt \citep{2002Icar..156..399B,2018Icar..312..181G} -- possibly in the Polana or Eulalia families like the targets of the asteroid sample return missions Hayabusa 2 and OSIRIS-REX \citep{2015Icar..247..191B}. The aim of our research is to establish where on the comet-asteroid continuum Wilson-Harrington really lies. Is it a traditional comet, sourced from beyond Neptune, that we caught the last gasps of activity from? Or does it have more in common with the active asteroids sourced from the Main Belt, possibly including Ryugu, Bennu, and Phaethon? We obtained new and re-analyzed archival observations of Wilson-Harrington, compared to meteorite samples, conducted new orbital integrations, and came to the following conclusions.

\begin{itemize}
    \item The reflectance spectrum of the nucleus of Wilson-Harrington as measured in the visible by LDT/Deveny \citep{2014SPIE.9147E..2NB} and IRTF/SpeX \citep{2003PASP..115..362R} in the near-infrared is unlike other comets, with an approximately neutral overall slope and a reflectance maximum near $\sim0.8-0.9\mu{m}$. It is well matched by laboratory spectra of the CM Chondrite Murchison or other carbonaceous chondrite meteorites. To our knowledge, this is the first time the nucleus of a designated comet has had a plausible meteoritical match, but if the object is asteroidal in origin then the importance of this is lessened.
    \item The linear phase curve coefficient of Wilson-Harrington's phase curve was estimated at $\beta_o=0.0427\pm0.0008$ and $\beta_c=0.0401\pm0.0006$ magnitudes per degree in the ATLAS \citep{2018PASP..130f4505T} `orange' and `cyan' filters, respectively. These values are within $1-\sigma$ of the coefficients derived for Wilson-Harrington over a narrower range of phase angles during a previous apparition \citep{2011ApJ...726..101I}. The median value for cometary nuclei is $\beta=0.042$ magnitudes per degree \citep{2017MNRAS.471.2974K}, while that of (101955) Bennu is $\beta=0.04$ \citep{2013Icar..226..663H}, thus Wilson-Harrington's phase curve is compatible with either affinity.
    \item Wilson-Harrington's recent orbital evolution is relatively stable with no significant close encounters. No periods of a sungrazing orbit that might have thermally altered its surface were seen before the orbit diverged though earlier warmer epochs thousands of years ago cannot be excluded by the present analysis. Wilson-Harrington is now and was in the 1970s and 1980s warmer than it was during its period of activity in 1949, and thus we argue against a purely thermal process being the driver of its single activity epoch.
\end{itemize}

While we cannot exclude a scenario whereby an unknown process has acted to change the surface of Wilson-Harrington from an originally cometary one to its modern one, we believe that a fundamentally asteroidal origin is more likely to explain its surface properties and orbit. However, if this is the case, there must be a way to maintain significant supplies of volatiles -- likely even hypervolatiles like carbon monoxide! -- on near-Earth objects if the \citet{1997Icar..128..114F} interpretation of its 1949 activity as being gas-dominated is correct. Inspection and analysis of the returned samples from Ryugu and Bennu -- and thus a constraint on their volatile budgets and histories -- might be critical to assessing how likely this scenario is and to establishing or refuting a chemical link between the bodies in addition the orbital similarities.

\begin{acknowledgments}
Lowell is a private, non-profit institution dedicated to astrophysical research and public appreciation of astronomy and operates the LDT in partnership with Boston University, the University of Maryland, the University of Toledo, Northern Arizona University and Yale University.

The authors wish to recognize and acknowledge the very significant cultural role and reverence that the summit of Maunakea has always had within the indigenous Hawaiian community. We are most fortunate to have the opportunity to conduct observations from this mountain

This work has made use of data from the Asteroid Terrestrial-impact Last Alert System (ATLAS) project. The Asteroid Terrestrial-impact Last Alert System (ATLAS) project is primarily funded to search for near earth asteroids through NASA grants NN12AR55G, 80NSSC18K0284, and 80NSSC18K1575; byproducts of the NEO search include images and catalogs from the survey area. This work was partially funded by Kepler/K2 grant J1944/80NSSC19K0112 and HST GO-15889, and STFC grants ST/T000198/1 and ST/S006109/1. The ATLAS science products have been made possible through the contributions of the University of Hawaii Institute for Astronomy, the Queen’s University Belfast, the Space Telescope Science Institute, the South African Astronomical Observatory, and The Millennium Institute of Astrophysics (MAS), Chile.

T. Kareta was supported in part by the Lowell Observatory Marcus Comet Research Fund.

Parts of this research was funded by NASA Solar System Observations grant 80NSSC20K0632 (PI: Reddy) and NASA Near-Earth Object Observation program grant NNX17AJ19G (PI: Reddy). 
\end{acknowledgments}

\vspace{5mm}
\facilities{IRTF(SpeX), LDT (DeVeny, LMI)}

\software{PhotometryPipeline \citep{2017A&C....18...47M}, emcee \citep{2013PASP..125..306F}, Rebound \citep{2012A&A...537A.128R}, PypeIt \citep{pypeit:joss_pub, pypeit:zenodo}}

%\bibliography{sample631}{}

\bibliographystyle{aasjournal}

\end{document}